# SEA LEVEL RISE BY CLIMATE CHANGE
**An order-of-magnitude approach to an environmental problem**[1]

Cedric Loretan[1,2], Andreas Müller[1]
1 Department of Physics and Institute of Teacher Education, University of Geneva, Pavillon d'Uni Mail (IUFE), Boulevard du Pont d'Arve 40, 1211 Geneva, Switzerland; Email: andreas.mueller@unige.ch
2 Lycée-Collège de l'Abbaye de St-Maurice, 1890 St-Maurice, Switzerland

## Question

Fig. 1, taken from a science literacy test, illustrates the distribution of water in different places on Earth, imagining that the total quantity is contained in 100 buckets [6]. Can you deduce from this figure, and other basic, publicly available geographic information the approximate value of about 70m for the rise in sea level by the melting of all land ice on Earth found in the media and the public debate [7]?

| | |
|---|---|
| About 2 buckets are filled with ice from glaciers and polar areas | 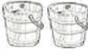 |
| Roughly 1½ buckets are filled with groundwater | 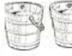 |
| Less than a bucket is filled with water from streams and lakes | 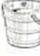 |
| Roughly ½ bucket is filled with the water in the atmosphere | 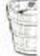 |
| And more than 95 buckets contain sea water | 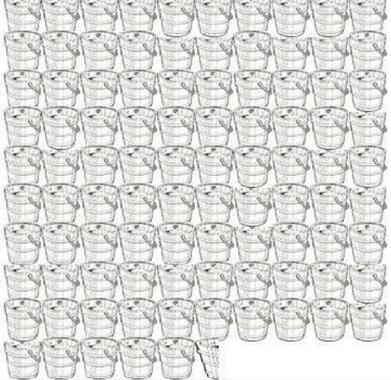 |

Fig. 1: Representation of how water is distributed on Earth

## 1    Answer

According to Fig. 1, the amount of water in glaciers (mainly in Greenland and the Antarctica) is equal to 2% of the amount of water in the seas (2 buckets / 95 buckets ≈ 0.021). A very simplified reasoning treats the oceans on Earth as a rectangular basin of fixed area, where adding x% of water will produce a rise of x% of the total depth. Now we need a value for the latter, and we will present three approaches for this:

(a) Simple estimate: If we look in an atlas, a typical depth of the oceans is a few kilometers, of which 2% gives several tens of meters of change.

(b) Elaborate estimate: In the case of a lower and upper limit given by an order-of-magnitude estimate, a good approximation is often given by their geometric mean [1, Ch. 1]. Choosing a less deep and a very deep part of the world ocean: North Sea, ~ 1 km (700m; Norwegian trench); and the Pacific Ocean ~ 10 km (11km; Marian trench) yields a geometric mean of ≈ 3 km, and thus an estimated sea level rise of 60 m.

(c) Publicly available sources: the average depth of the world ocean is known to be 3700 m [8], so 2% correspond roughly to 70 m. Note this tabulated average ocean depth coincides with the estimated depth of (b) within 20%, and the same then holds for the sea level rise estimations based on these depth values.

---

[1] A short version of this paper has appeared as Loretan, C., Müller, A. (2022). A drop in the bucket? In: Adams, J. (series edt.), Fermi Questions. *The Physics Teacher*, 60 (Dec.) A796

## 2   Comments, further considerations

<u>Literature values</u>: An illustrative and well established way of discussing various contributions to the sea level (change) is the "see-level equivalent" (SLE), i.e., the change in global average sea level that would occur if a given amount of ice (or water) was added to (or removed from) the world oceans. For Greenland and Antarctica (which contain 99.5m of all land ice), the SLE is 7.4 m and 58.3 m, respectively [2], i.e. a total of 65.7 m, with which the above estimated are well in accord. Note that to this value 1-2 m have to be added due to thermal expansion [3].

<u>Shape considerations</u>: First, while we assumed above a rectangular basin for simplicity, it shape actually does not matter, as long as the walls are vertical: For container with arbitrary, but constant cross-section $A$ the ("filling") volume $V = A \cdot h$ and ("filling") height $h$ are proportional, as assumed above. Second, the question of inclined walls needs more consideration. Our oceanic "basin" indeed has rather shallow parts, called "continental shelves". They extend in most cases at most up to a few hundred kilometers (average ≈ 80km), which is an order of magnitude smaller the dimension of the oceans. This implies that out estimated value is an overestimation, but will not change its order-of-magnitude.

<u>Educational rationale</u>: Note that geometric simplifications of the kind made here belong to essential "toolkit" of order-of-magnitude reasoning, as nicely illustrated by the book title "Consider a Spherical Cow" [4]. Similarly, exploiting different ways of getting input data and checking for consistency are also part of this toolkit.

Specifically, the purpose of the present example is to show, how easily available sources found in daily life (school science test, maps) may, when combined with order-of-magnitude reasoning, allow to draw important conclusions. In all cases considered above, students can find the right order of magnitude for a value used in a discussion of high societal importance, with their own reasoning, and not just by believing values found in the media, We consider this as an exercise in critical thinking, as important component of science education [5].

## References and Links